\documentclass[letter, twocolumn]{jpsj/jpsj3}

\usepackage{mathtools}
\usepackage{color}
\usepackage{subfigure}
\usepackage[T1]{fontenc}

\begin{document}

\title{Graph minor embedding can affect sampling degenerate ground states using quantum annealing}

\author{%
Naoki Maruyama$^{1,4}$, Masayuki Ohzeki$^{1,2,3,4}$ and Kazuyuki Tanaka$^{1}$
}
\inst{%
$^1$Graduate School of Information Science, Tohoku University, Sendai 980-8579, Japan,
$^2$Department of Physics, Institute of Science Tokyo, Tokyo 152-8551, Japan,
$^3$Semiconductors and Informatics, Kumamoto University, Kumamoto 860-8555, Japan,
$^4$Sigma-i Co., Ltd., Tokyo 108-0075, Japan
}

\abst{%
Quantum annealing, as currently implemented in hardware, cannot fairly sample all ground states.
Graph minor embedding, which maps a problem to the hardware graph of quantum annealers, affects sampling all states.
In this study, we demonstrate the influence of graph minor embedding on fair sampling of degenerate ground states.
For two embedded models that introduce auxiliary variables, numerical simulations of Schrödinger evolution revealed that fairness varies significantly depending on the embedding, and the chain strength is related to ground-state fairness.
Using perturbation theory, we found that chain strength determines the energy landscape around ground states, with flatter landscapes having higher probabilities of being obtained.
}

\maketitle

\textit{Introduction.} Quantum annealing (QA) ~\cite{kadowaki1998} is a meta-heuristic for solving combinatorial optimization problems using quantum effects.
Since the development of hardware that implements QA, such as D-Wave quantum annealer ~\cite{johnson2010,harris2010}, QA applications in various fields have been studied ~\cite{neukart2017,ohzeki2018a,ide2020,arai2020,rosenberg2015,venturelli2016,stollenwerk2020a}.
Compared to research focused on obtaining optimal solutions to optimization problems, few studies focus on sampling diverse solutions.

In standard QA, it is not always possible to obtain all ground states with equal probability ~\cite{matsuda2009}.
Recent versions of quantum annealers have also been reported to sample all solutions unfairly ~\cite{mandra2017}.
Although it is known that fair sampling can be achieved by using higher-order driver Hamiltonians ~\cite{matsuda2009,konz2019}, implementing such complex driver Hamiltonians in hardware is currently challenging.
In applications that use multiple solutions, such as SAT filters ~\cite{weaver2012,azinovic2017} and machine learning ~\cite{hinton2002,eslami2014}, it is desirable to sample states without bias.
For the application of QA, it is crucial to understand the mechanism of unfair sampling.

Quantum annealer needs to transform specific problems to treat them on the hardware, one of which is graph minor embedding.
This process maps logical variables to physical qubits on the hardware graph.
It is challenging to express an original problem directly, although the connectivity of D-Wave quantum annealers has gradually increased ~\cite{dattani2019,boothby2020}.
Then one can add auxiliary variables and ferromagnetic interactions called \textit{chain}.
For optimization purposes, chain strength is involved with the performance by QA ~\cite{choi2008}.

The effect of graph minor embedding appears more pronounced in sampling than in optimization.
While optimization aims to achieve a single ground state, sampling must consider all possible states.
Graph minor embedding maintains the ground states but ignores other states, thus having a nontrivial impact on sampling results.
It is known that, depending on the problem size and embedding size, the probability that samples exist in the logical subspace (without chain breaks) is exponentially suppressed ~\cite{marshall2020}.
A study that extended this research on thermal sampling to quantum sampling ~\cite{marshall2021} shows that embedding shifts the phase transition boundary.

In this study, we demonstrate the influence of graph minor embedding on fair sampling of degenerate ground states.
For problems that do not require auxiliary variables, it has been demonstrated that quantum annealers cannot fairly sample ground states ~\cite{mandra2017}.
We present the influence of embedding in problems that require auxiliary variables.
For two types of embedded models with different numbers of auxiliary variables, numerical solution by the Schrödinger equation revealed that the fairness of ground states varies significantly depending on the embedding method.
We also show that the chain strength introduced by graph minor embedding is related to the fairness of ground states.
Using perturbation theory to investigate this mechanism, we found that the magnitude of chain strength determines the energy landscape around the ground states, and states with flatter energy landscapes tend to have higher probabilities of being obtained.

\textit{Methods.} A Hamiltonian of transverse-field Ising model with $N$ spins is defined by,
\begin{equation}
H(t)=-\left(1-\frac{t}{\tau}\right) \sum_{i} \sigma_{i}^{x}+\frac{t}{\tau} H_{0}(\{\sigma_i^{z}\}),
\end{equation}
where $H_{0}$ is a target Hamiltonian, and $\sigma_i^{x}, \sigma_i^{z}$ are $x$ and $z$ component of the Pauli operator acting on site $i$.
$H(t)$ changes from $-\sum_i \sigma_{i}^{x}$ at $t=0$ to $H_{0}$ at $t=\tau$.
If the process is adiabatic, the system is more likely to follow the instantaneous ground state, leading to a nontrivial final ground state of $H_{0}$ ~\cite{suzuki2005}.

\begin{figure}[tbh]
\centering
\subfigure[]{\includegraphics[width=0.15\textwidth]{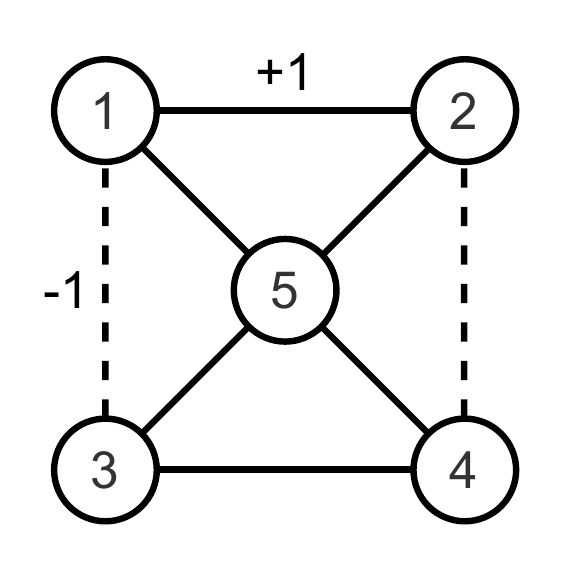}\label{fig:5spin}}
\subfigure[]{\includegraphics[width=0.15\textwidth]{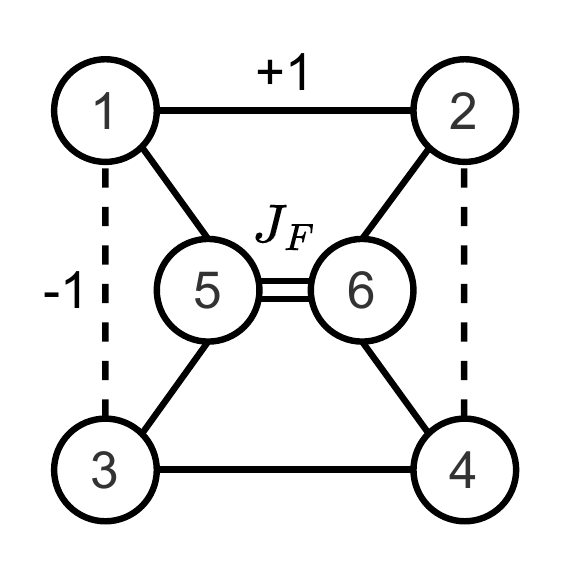}\label{fig:5spin_embed_centralchain}}
\subfigure[]{\includegraphics[width=0.15\textwidth]{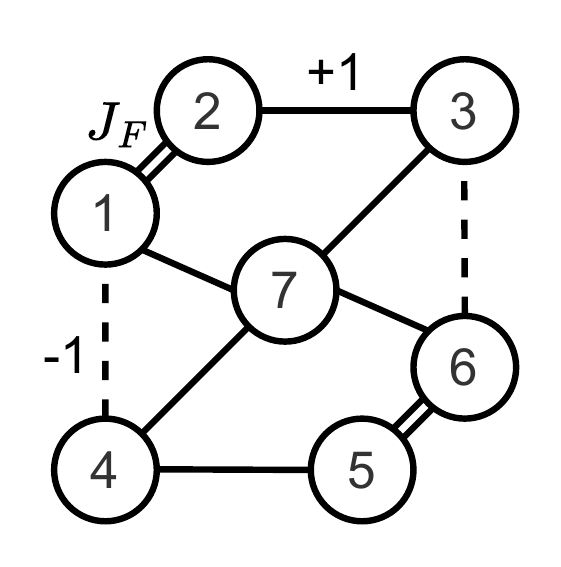}\label{fig:5spin_embed_edgechain}}
\caption{(a) Original model, (b) Embedded model with the central chain (minimal embedding), and (c) Embedded model with the edge chains (redundant embedding). The solid and dashed lines represent ferromagnetic interactions $J_{ij}=+1$ and anti-ferromagnetic interactions $J_{ij}=-1$. The double line represents a chain $J_F$.}
\end{figure}

We assume a five-spin system ~\cite{matsuda2009,konz2019} as shown in Figure \ref{fig:5spin},
whose target Hamiltonian is given by $H_{0} = -\sum_{\langle i j\rangle} J_{i j} \sigma_{i}^{z} \sigma_{j}^{z}$, where all interactions $J_{ij} = \pm 1$.
This system has six ground states: $|\uparrow \uparrow \uparrow \uparrow \uparrow\rangle,|\uparrow \uparrow \downarrow \downarrow \uparrow\rangle,|\uparrow \uparrow \downarrow \downarrow \downarrow\rangle$, and the states in which all the spins are inverted.
We denote these states as $|1\rangle$, $|2\rangle$, and $|3\rangle$.
As this model also has spin-inversion symmetric ground states, with equal probabilities for the two spin-symmetric states, we will focus on the three states mentioned above.
We consider two embedded models with different embedding configurations.
The first model (Figure \ref{fig:5spin_embed_centralchain}) uses the central chain configuration, embedding the fifth spin using a single central chain.
This configuration has the minimum number of additional spins required by the Pegasus graph of D-Wave Advantage ~\cite{dattani2019,boothby2020}.
The second model (Figure \ref{fig:5spin_embed_edgechain}) adopts the edge chains configuration, embedding the first and fourth spins redundantly using edge chains.
Hereafter, for brevity, we refer to these as the minimally embedded and redundantly embedded models, respectively.

\textit{Results.} For the three types of models explained in the previous section, we investigate the effects of embedding in standard QA by directly solving the time-dependent Schrödinger equation.
We evaluate the fairness of the sampling results using a metric employed in a previous study ~\cite{kadowaki2019}.
We divide the three ground states into two groups: $S = \{ |1\rangle \}$ and $C = \{ |2\rangle, |3\rangle \}$ because the states in $C$ have inversion symmetry and their probabilities are equal, as later experimental results show.
The probabilities of the states in the two groups are defined as follows:
\begin{equation}
P_G(t) = \frac{1}{|G|} \sum_{j \in G}|\langle\psi(t) \mid j\rangle|^2,
\end{equation}
where $G \in \{ S, C \}$, $|\psi(t)\rangle$ is the state of the system at time $t$, and $|G|$ is the number of states in group $G$.
We use the ratio of probabilities $P_S/P_C$ as a metric of fairness.
The closer $P_S/P_C$ is to 1, the fairer it means.

\begin{figure}[tbh]
\centering
\subfigure[The minimally embedded model]{\includegraphics[width=\linewidth]{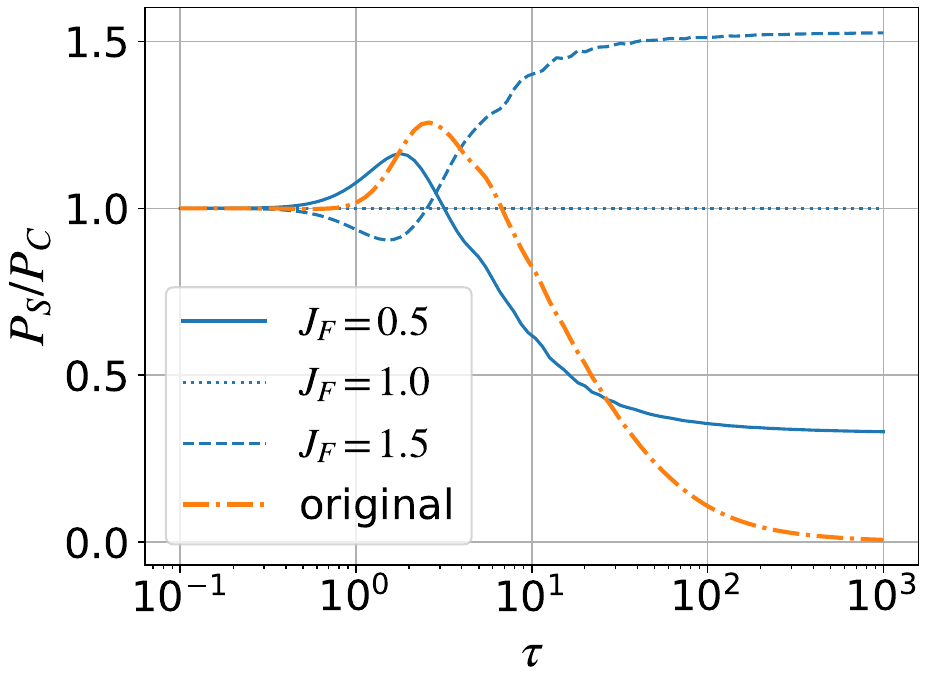}\label{fig:tau_vs_prob_ratio_centralchain}}
\subfigure[The redundantly embedded model]{\includegraphics[width=\linewidth]{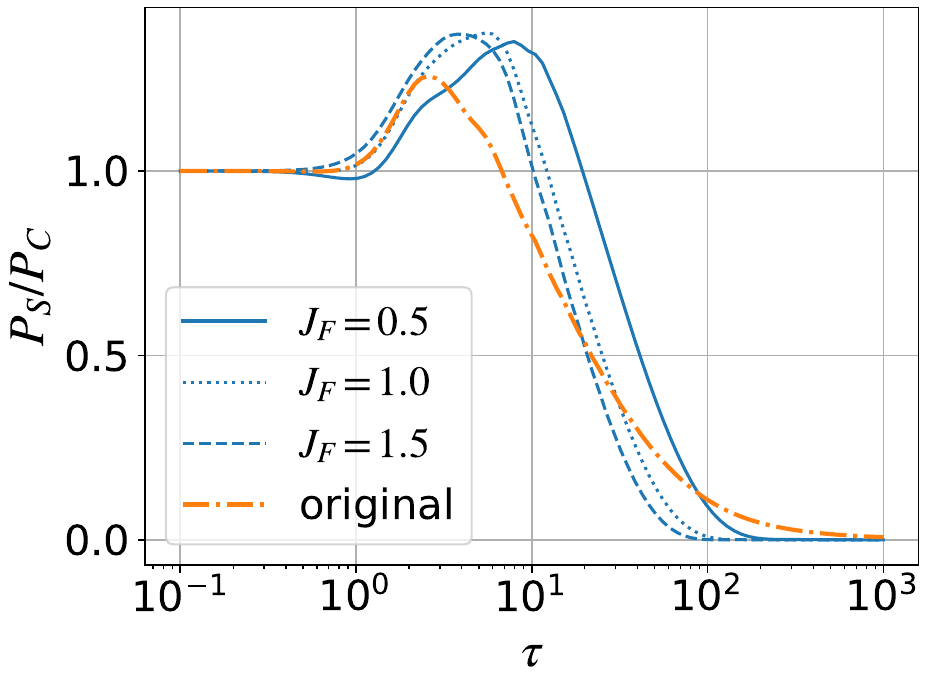}\label{fig:tau_vs_prob_ratio_edgechain}}
\caption{Annealing time dependence on the ratio of probabilities in the original and minimally embedded model.
The solid, dotted, and dashed lines represent $J_F \in \{0.5, 1.0, 1.5\}$ in the embedded model, and the dash-dot line is in the original model.}
\end{figure}

We show the dependence of annealing time $\tau$ on the ratio for $J_F \in \{0.5, 1.0, 1.5\}$ in the two types of embedded models.
In the results of the minimally embedded model in Figure \ref{fig:tau_vs_prob_ratio_centralchain}, for large $\tau$, the ratio is approaching zero in the original model. Still, the behavior differs in the embedded model.
The dependence on chain strength can be confirmed.
In particular, all the ground states can be obtained fairly when $J_{F}=1$.
On the other hand, as shown in Figure \ref{fig:tau_vs_prob_ratio_edgechain}, in the redundantly embedded model, for large $\tau$, the ratio converges to zero both before and after embedding.
Regarding the dependence on chain strength, although differences in the ratio are observed only in the small annealing time region ($\tau < 10^2$), they are not observed in the larger area.

From the numerical results, we found that the chain strength is related to the fairness of sampling ground states, depending on the embeddings.
Following previous studies ~\cite{sieberer2018,konz2019}, we employ perturbation theory to elucidate this mechanism.
If we anneal adiabatically, i.e., $\tau$ large enough, the instantaneous ground states are exponentially suppressed, which means towards the end of annealing at $\tau - \lambda$ (for a small $\lambda > 0$) the system is in the ground state of $H\left( \tau - \lambda \right)$.
This observation is crucial for predicting the probabilities of different degenerate ground states.
These probabilities are given by squaring the amplitudes of the lowest eigenvector of $H\left(\tau - \lambda\right)$, assuming that the driver's small contribution lifts the degeneracies.
Because $H\left(\tau - \lambda\right)$ can be viewed as $H_{0}$ perturbed by $H_{1}$, we analyze fair sampling using a perturbative approach.
We define the Hamiltonian around the final time as $H(\lambda) = H_{0} + \lambda V$,
where $\lambda > 0$ is a sufficiently small coefficient and $V$ is a driver Hamiltonian, which is recognized as a perturbation for the target Hamiltonian $H_{0}$.
As mentioned above, we focus on transverse-field $V = -\sum_{i=1}^{N} \sigma_{i}^{x}$.

Given $d$ ground states $|n^{(0)}\rangle$ of $H_{0}$ with energy $E_{n}^{0}$, the eigenstate $|n(\lambda)\rangle$ and eigenenergy $E_{n}^{\lambda}$ of $H(\lambda)$ are expanded as $|n(\lambda)\rangle=\left|n^{(0)}\right\rangle+\lambda\left|n^{(1)}\right\rangle+\lambda^{2}\left|n^{(2)}\right\rangle+\ldots$, $E_{n}(\lambda)=E_{n}^{(0)}+\lambda E_{n}^{(1)}+\lambda^{2} E_{n}^{(2)}+\ldots$ for small $\lambda$.
Let $P_{1}$ be a project operator onto the first-level degenerate eigenspace of $H_{0}$.
The first-order perturbation satisfies the following equation in the degenerate level,
\begin{equation}
\label{eq:1st_perturb}
E_{n}^{(1)}\left|n^{(0)}\right\rangle=P_{1} V P_{1} \left|n^{(0)}\right\rangle,
\end{equation}
where the first-order energy correction $E_{n}^{(1)}$ and the zero-order state correction $\left|n^{(0)}\right\rangle$.

In the original model and embedded models shown in Figure \ref{fig:5spin}, first-order perturbation solves the degeneracy except for the degeneracy that cannot be solved due to spin-inversion symmetry.
For example, when proceeding with calculations by removing spin-inversion symmetry, a small longitudinal magnetic field can be applied to the first spin.
On the other hand, in the embedded model shown in Figure \ref{fig:5spin_embed_centralchain}, first-order perturbation does not solve the degeneracy of the ground states.
That is, $\left\langle n^{(0)}\left|P_1 V P_1\right| n^{(0)}\right\rangle=0$.
State $|1\rangle$ is not obtained in the original model (a) but more in the embedded model (b).
In the first-order perturbation, the eigenvalue of state $|1\rangle$ is zero and lower than the eigenvalues of states $|2\rangle$ and $|3\rangle$ in the original model (a), while the eigenvalue of state $|1\rangle$ is the same as the eigenvalues of states $|2\rangle$ and $|3\rangle$ in the embedded model (b).
Therefore, state $|1\rangle$ is obtained more in the embedded model (b) than in the original model (a).
Also, in the embedded model shown in Figure \ref{fig:5spin_embed_edgechain}, like the original model (a), first-order perturbation solves the degeneracy except for the degeneracy that cannot be solved due to spin-inversion symmetry.
Therefore, the redundantly embedded model shows the same unfairness in the large annealing time as the original model.

When the first-order perturbation degenerates, the second-order perturbation must be considered.
Let $P_{2}$ be a project operator to the degenerate subspace of $P_{1}VP_{1}$.
The second-order perturbation satisfies the following equation in the degenerate level,
\begin{equation}
E_{n}^{(2)}\left|n^{(0)}\right\rangle=P_{2} W P_{2}\left|n^{(0)}\right\rangle,
\label{eq:2nd_perturb}
\end{equation}
where $W = VQ (E_{n}^{(0)}-H_{0})^{-1} Q V$, and $Q=1-P_{1}$.

\begin{figure}[tbh]
\begin{center}
\includegraphics[width=\linewidth]{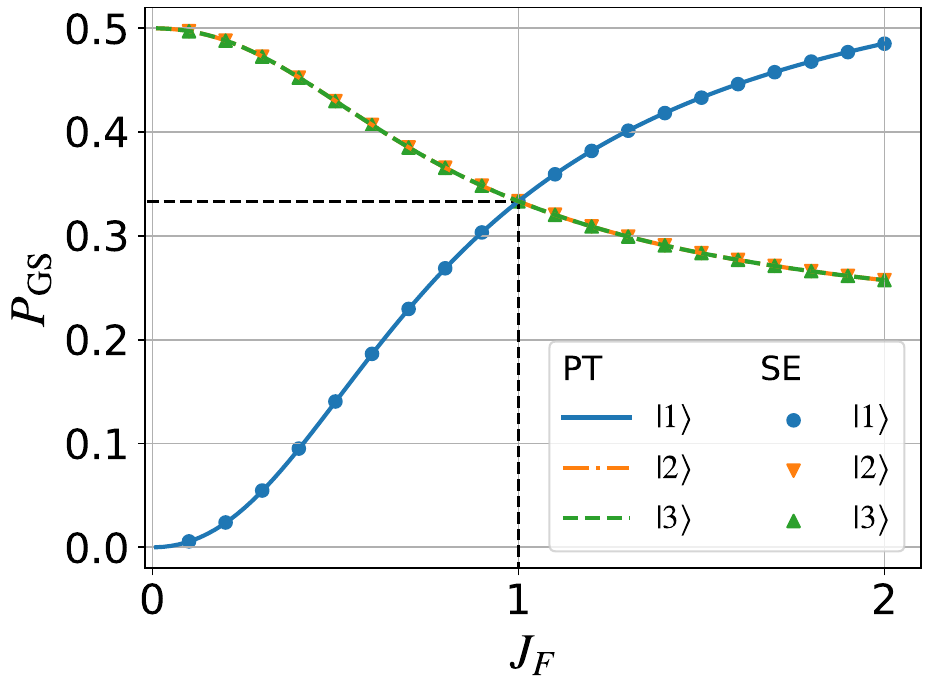}
\caption{
Chain strength dependence on the probability of obtaining ground states in $\tau=1000$ by the perturbation theory (PT) and the Schr\"{o}dinger equation (SE).
Note that the lines (PT) and points (SE) for $\left|2\right\rangle$ and $\left|3\right\rangle$ overlap.
}
\label{fig:chain_strength_vs_prob}
\end{center}
\end{figure}

Solving this equation attains the second-order energy correction $E_{n}^{(2)}$ and the zero-order state correction $\left|n^{(0)}\right\rangle$.
The probability of obtaining ground states at the final time can be calculated from the eigenvector for the minimal eigenvalue of $P_{2}WP_{2}$.
Figure \ref{fig:chain_strength_vs_prob} shows the probability of obtaining ground states obtained by perturbation theory concerning chain strength.
The direct solution of the Schrödinger equation at $\tau=1000$ is also shown, and we confirm that both agree.

Next, we investigate how chain strength relates to the probability of obtaining ground states.
We denote the matrix $-P_2 W P_{2}$ as $A^{(2)}$, in the embedded model shown in Figure \ref{fig:5spin_embed_centralchain}, which is expressed as follows:
\begin{equation}
\label{eq:adj_mat}
\begin{split}
&A^{(2)}\\
&=\begin{bmatrix}
\frac{2 J_{F}+5}{J_{F}+2} & 1 & 0 & 0 & 0 & 1 \\
1 & \frac{4J_{F}+3}{3J_{F}} & \frac{1}{J_{F}} & 0 & 0 & 0 \\
0 & \frac{1}{J_{F}} & \frac{4J_{F}+3}{3J_{F}} & 1 & 0 & 0 \\
0 & 0 & 1 & \frac{2 J_{F}+5}{J_{F}+2} & 1 & 0 \\
0 & 0 & 0 & 1 & \frac{4J_{F}+3}{3J_{F}} & \frac{1}{J_{F}} \\
1 & 0 & 0 & 0 & \frac{1}{J_{F}} & \frac{4J_{F}+3}{3J_{F}}
\end{bmatrix}.
\end{split}
\end{equation}
Specifically, the components of $A^{(2)}$ can be expressed as
\begin{equation}
A_{ii}^{(2)} = \sum_{j \in \partial i} \left( e_{j} - e_{i} \right)^{-1}, \ A_{ij}^{(2)} = \sum_{k \in \partial (i, j)} \left( e_{k} - e_{i} \right)^{-1}.
\label{eq:2nd_perturb_matrix}
\end{equation}
Here, $\partial i$ is the set of states that can be reached by a single spin flip from ground state $i$, $\partial (i, j)$ is the set of states that are passed through when transitioning from ground state $i$ to other ground state $j$ by two spin flips, and $e_k$ is the energy of state $k$ (including excited states).
From this equation, we can understand that each component represents the energy gap around the ground state.
The diagonal components $A_{ii}^{(2)}$ are the reciprocals of energy differences with states with one spin flip from the ground state.
For example, the (1,1)-component is the sum of reciprocals of energies of neighboring states of state $|1\rangle=|\uparrow\uparrow\uparrow\uparrow\uparrow\uparrow\rangle$: $|\uparrow\uparrow\uparrow\uparrow\uparrow\downarrow\rangle, |\uparrow\uparrow\uparrow\uparrow\downarrow\uparrow\rangle, \cdots, |\downarrow\uparrow\uparrow\uparrow\uparrow\uparrow\rangle$.
The off-diagonal components $A_{ij}^{(2)}$ represent the reciprocals of energy barrier magnitudes during transitions from one ground state to another.
For example, the (2,3)-component is the sum of reciprocals of energies of excited states $|\uparrow\uparrow\downarrow\downarrow\uparrow\downarrow\rangle,|\uparrow\uparrow\downarrow\downarrow\downarrow\uparrow\rangle$ that are passed through when transitioning from state $|2\rangle=|\uparrow\uparrow\downarrow\downarrow\uparrow\uparrow\rangle$ to $|3\rangle=|\uparrow\uparrow\downarrow\downarrow\downarrow\downarrow\rangle$.
The above equation shows that these components, i.e., the energy landscape around the ground states, change according to the chain strength $J_{F}$.
To numerically verify the relationship between energy landscape and ground state occurrence probability, we define the relative flatness of the energy landscape of ground state $i$ as
\begin{equation}
\mathrm{REF}_i \coloneqq \frac{\sum_{k \neq i} A_{ik}^{(2)}}{\sum_{j \neq i} \sum_{k \neq j} A_{jk}^{(2)}}.
\end{equation}
This metric is the sum of reciprocals of energy barrier magnitudes during transitions from ground state $i$ to other ground states $j$, contained in the off-diagonal components $A_{ij}^{(2)}$.
The smaller the energy differences between other states, the larger the value of this index, meaning that the energy landscape is flatter.

\begin{figure}[tbh]
\begin{center}
\includegraphics[width=\linewidth]{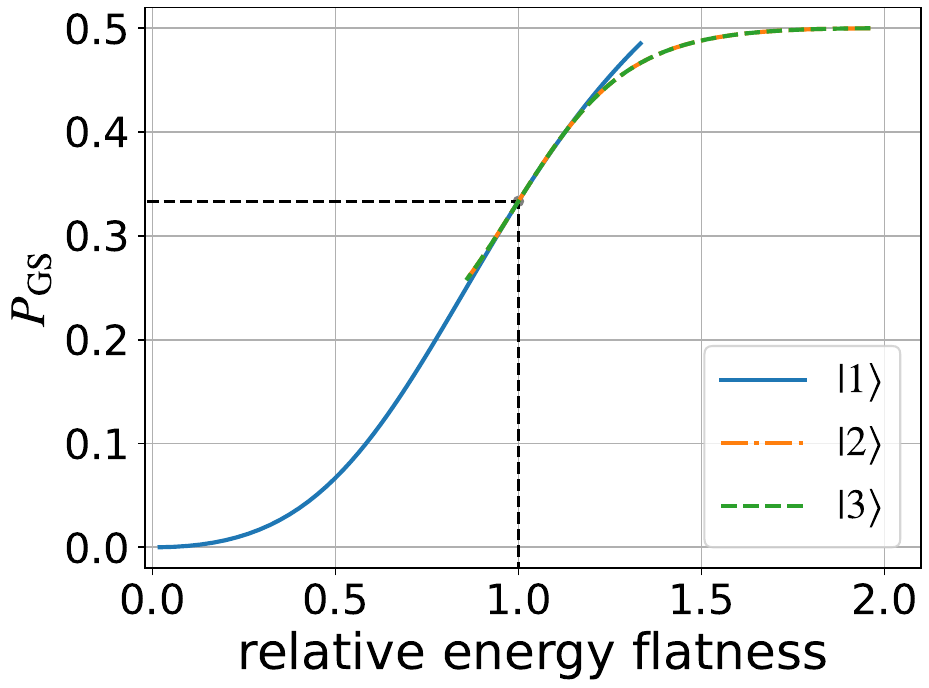}
\caption{
Relationship between the probability of obtaining ground states and the relative flatness of their energy landscapes.
Note that the lines $\left|2\right\rangle$ and $\left|3\right\rangle$ overlap.
}
\label{fig:flatness_vs_prob}
\end{center}
\end{figure}

Figure \ref{fig:flatness_vs_prob} shows the relationship between the relative flatness of the energy landscape and the probability of obtaining ground states for chain strength $J_F \in (0,2)$.
As can be seen from this figure, the flatter the energy landscape, the higher the probability of obtaining that state.
In particular, when the flatness of the energy landscape equals 1.0 ($J_F=1.0$), the probability of obtaining each ground state becomes equal to 1/3, thereby achieving fair sampling.

\textit{Discussion.} We have demonstrated experimentally and analytically that graph minor embedding influences sampling degenerate ground states by QA.
Our results show that the probability distribution of ground states in the embedded model differs from that in the original model.
We employ perturbation theory to analyze the probabilities of ground states and verify that the analysis aligns with our numerical results.
Our analysis also reveals that the chain strength introduced by graph minor embedding affects the energy landscape around ground states, thereby influencing the probabilities of obtaining ground states.
The relationship between flatness and fairness should also hold in larger-scale problems.
In this case, the distance between ground states increases, so for simple drivers like a transverse field, the influence of the energy landscape between ground states becomes more significant, as seen in the second-order correction.
While our toy problem is embedded into the Pegasus graph of D-Wave Advantage ~\cite{dattani2019,boothby2020}, our argument holds regardless of the type of hardware graph, such as the Chimera graph and the Zephyr graph ~\cite{pelofske2023}.

We also found the relationship between the probability of obtaining ground states and the flatness of their energy landscapes.
The flatter the energy landscape, the higher the frequency of obtaining that state.
Such a relationship has been reported with similar results in the past study ~\cite{matsuda2009}, where free spins (spins that can be flipped without incurring an energy cost) are interpreted as the flatness of the energy landscape.
Previous studies ~\cite{baldassi2018,ohzeki2018} have mentioned that QA favors flat minima because the larger local entropy of densely clustered configurations lowers their effective free energy, making these states statistically more likely to emerge.
Figure 3 in the study ~\cite{konz2019} shows that the parameter $J_{3,4}$ in the toy problem can arbitrarily change the sampling bias, which is similar to our embedded models.
Unlike these studies, our results link flatness to quantitatively controllable hardware elements, such as embeddings and chains.
Graph minor embedding is usually recognized as a bottleneck that reduces the accuracy of optimization and sampling by quantum annealers.
On the other hand, we expect that the embedding process will be available for mitigating the unfairness of ground states, similar to quantum annealing corrections that improve optimization performance~\cite{vinci2015,bauza2024}.
Note that when evaluating flatness rigorously, the diagonal components we ignored this time, i.e., energy differences between ground states, should be included in the definition.

As the results regarding the embedded model in Figure \ref{fig:5spin_embed_edgechain} show, why do embedding processes not necessarily affect sampling fairness?
We expect this to be related to the graph structure of the ground states in the embedded models.
In the analysis using first-order perturbation, $-P_1 V P_1 =: A^{(1)}$ in the above equation represents the adjacency matrix of the graph of the ground states, which in turn means the graph structure of the ground states.
In second-order perturbation, we obtain the matrix $A^{(2)}$ that not only represents adjacency information but also includes chain strength.
By examining the graph structure of ground states in more detail, it may be possible to gain a deeper understanding of sampling fairness by QA.
For example, while it remains unclear how problems should be embedded to reduce embedding's influence, we anticipate that the analysis using the adjacency matrix will shed light on this critical question.

The results of this study help consider the differences between quantum annealer results and theoretical results.
On the other hand, it is necessary to consider influences other than graph minor embedding, such as the effects of external thermal baths ~\cite{kadowaki2019} and control errors ~\cite{chancellor2020}.

\textit{Acknowledgement}.
We thank Masayuki Yamamoto and Manaka Okuyama for the fruitful discussion.
This study was financially supported by programs for bridging the gap between R\&D and IDeal society (Society 5.0) and Generating Economic and social value (BRIDGE) and Cross-ministerial Strategic Innovation Promotion Program (SIP) from the Cabinet Office 23836436.

\textit{Author contributions}.
N.M. conceived of the presented idea and performed the experiments.
M.O. supervised the findings of this work.
K.T. helped to supervise the project.
All authors discussed the results and contributed to the preparation of the final manuscript.

\bibliographystyle{jpsj/jpsj}
\bibliography{main}

@misc{arai2020,
  title = {Mean Field Analysis of Reverse Annealing for Code-Division Multiple-Access Multiuser Detection},
  author = {Arai, Shunta and Ohzeki, Masayuki and Tanaka, Kazuyuki},
  year = 2020,
  month = apr,
  journal = {arXiv.org},
  doi = {10.1103/PhysRevResearch.3.033006},
  urldate = {2024-07-23},
  howpublished = {https://arxiv.org/abs/2004.11066v4},
  langid = {english}
}

@article{azinovic2017,
  title = {Assessment of {{Quantum Annealing}} for the {{Construction}} of {{Satisfiability Filters}}},
  author = {Azinovi{\'c}, Marlon and Herr, Daniel and Heim, Bettina and Brown, Ethan and Troyer, Matthias},
  year = 2017,
  month = apr,
  journal = {SciPost Physics},
  volume = {2},
  number = {2},
  pages = {013},
  issn = {2542-4653},
  doi = {10.21468/SciPostPhys.2.2.013},
  urldate = {2021-03-07},
  langid = {english}
}

@article{baldassi2018,
  title = {Efficiency of Quantum vs. Classical Annealing in Nonconvex Learning Problems},
  author = {Baldassi, Carlo and Zecchina, Riccardo},
  year = 2018,
  month = feb,
  journal = {Proceedings of the National Academy of Sciences},
  volume = {115},
  number = {7},
  pages = {1457--1462},
  issn = {0027-8424, 1091-6490},
  doi = {10.1073/pnas.1711456115},
  urldate = {2021-05-15},
  langid = {english}
}

@misc{bauza2024,
  title = {Scaling {{Advantage}} in {{Approximate Optimization}} with {{Quantum Annealing}}},
  author = {Bauza, Humberto Munoz and Lidar, Daniel A.},
  year = 2024,
  month = jan,
  number = {arXiv:2401.07184},
  eprint = {2401.07184},
  primaryclass = {cond-mat, physics:quant-ph},
  publisher = {arXiv},
  doi = {10.48550/arXiv.2401.07184},
  urldate = {2024-02-02},
  archiveprefix = {arXiv}
}

@article{boothby2020,
  title = {Next-{{Generation Topology}} of {{D-Wave Quantum Processors}}},
  author = {Boothby, Kelly and Bunyk, Paul and Raymond, Jack and Roy, Aidan},
  year = 2020,
  month = feb,
  journal = {arXiv:2003.00133 [quant-ph]},
  eprint = {2003.00133},
  primaryclass = {quant-ph},
  urldate = {2021-08-05},
  archiveprefix = {arXiv}
}

@article{chancellor2020,
  title = {Disorder-Induced Entropic Potential in a Flux Qubit Quantum Annealer},
  author = {Chancellor, Nicholas and Crowley, Philip J. D. and {\DJ}uri{\'c}, Tanja and Vinci, Walter and Amin, Mohammad H. and Green, Andrew G. and Warburton, Paul A. and Aeppli, Gabriel},
  year = 2020,
  month = jun,
  journal = {arXiv:2006.07685 [quant-ph]},
  eprint = {2006.07685},
  primaryclass = {quant-ph},
  urldate = {2021-02-02},
  archiveprefix = {arXiv}
}

@article{choi2008,
  title = {Minor-Embedding in Adiabatic Quantum Computation: {{I}}. {{The}} Parameter Setting Problem},
  shorttitle = {Minor-Embedding in Adiabatic Quantum Computation},
  author = {Choi, Vicky},
  year = 2008,
  month = oct,
  journal = {Quantum Information Processing},
  volume = {7},
  number = {5},
  pages = {193--209},
  issn = {1573-1332},
  doi = {10.1007/s11128-008-0082-9},
  urldate = {2023-09-16},
  langid = {english}
}

@article{dattani2019,
  title = {Pegasus: {{The}} Second Connectivity Graph for Large-Scale Quantum Annealing Hardware},
  shorttitle = {Pegasus},
  author = {Dattani, Nike and Szalay, Szilard and Chancellor, Nick},
  year = 2019,
  month = jan,
  journal = {arXiv:1901.07636 [quant-ph]},
  eprint = {1901.07636},
  primaryclass = {quant-ph},
  urldate = {2020-02-23},
  archiveprefix = {arXiv}
}

@article{eslami2014,
  title = {The {{Shape Boltzmann Machine}}: {{A Strong Model}} of {{Object Shape}}},
  shorttitle = {The {{Shape Boltzmann Machine}}},
  author = {Eslami, S. M. Ali and Heess, Nicolas and Williams, Christopher K. I. and Winn, John},
  year = 2014,
  month = apr,
  journal = {International Journal of Computer Vision},
  volume = {107},
  number = {2},
  pages = {155--176},
  issn = {1573-1405},
  doi = {10.1007/s11263-013-0669-1},
  urldate = {2021-03-07},
  langid = {english}
}

@article{harris2010,
  title = {Experimental Investigation of an Eight-Qubit Unit Cell in a Superconducting Optimization Processor},
  author = {Harris, R. and Johnson, M. W. and Lanting, T. and Berkley, A. J. and Johansson, J. and Bunyk, P. and Tolkacheva, E. and Ladizinsky, E. and Ladizinsky, N. and Oh, T. and Cioata, F. and Perminov, I. and Spear, P. and Enderud, C. and Rich, C. and Uchaikin, S. and Thom, M. C. and Chapple, E. M. and Wang, J. and Wilson, B. and Amin, M. H. S. and Dickson, N. and Karimi, K. and Macready, B. and Truncik, C. J. S. and Rose, G.},
  year = 2010,
  month = jul,
  journal = {Physical Review B},
  volume = {82},
  number = {2},
  pages = {024511},
  publisher = {American Physical Society},
  doi = {10.1103/PhysRevB.82.024511},
  urldate = {2021-03-23}
}

@article{hinton2002,
  title = {Training {{Products}} of {{Experts}} by {{Minimizing Contrastive Divergence}}},
  author = {Hinton, Geoffrey E.},
  year = 2002,
  month = aug,
  journal = {Neural Computation},
  volume = {14},
  number = {8},
  pages = {1771--1800},
  issn = {0899-7667},
  doi = {10.1162/089976602760128018},
  urldate = {2023-09-16}
}

@inproceedings{ide2020,
  title = {Maximum {{Likelihood Channel Decoding}} with {{Quantum Annealing Machine}}},
  booktitle = {2020 {{International Symposium}} on {{Information Theory}} and {{Its Applications}} ({{ISITA}})},
  author = {Ide, Naoki and Asayama, Tetsuya and Ueno, Hiroshi and Ohzeki, Masayuki},
  year = 2020,
  month = oct,
  pages = {91--95},
  issn = {2689-5854},
  urldate = {2025-08-05}
}

@article{johnson2010,
  title = {A Scalable Control System for a Superconducting Adiabatic Quantum Optimization Processor},
  author = {Johnson, M. W. and Bunyk, P. and Maibaum, F. and Tolkacheva, E. and Berkley, A. J. and Chapple, E. M. and Harris, R. and Johansson, J. and Lanting, T. and Perminov, I. and Ladizinsky, E. and Oh, T. and Rose, G.},
  year = 2010,
  month = apr,
  journal = {Superconductor Science and Technology},
  volume = {23},
  number = {6},
  pages = {065004},
  publisher = {IOP Publishing},
  issn = {0953-2048},
  doi = {10.1088/0953-2048/23/6/065004},
  urldate = {2021-03-23},
  langid = {english}
}

@article{kadowaki1998,
  title = {Quantum Annealing in the Transverse {{Ising}} Model},
  author = {Kadowaki, Tadashi and Nishimori, Hidetoshi},
  year = 1998,
  month = nov,
  journal = {Physical Review E},
  volume = {58},
  number = {5},
  pages = {5355--5363},
  publisher = {American Physical Society},
  doi = {10.1103/PhysRevE.58.5355},
  urldate = {2023-09-16}
}

@article{kadowaki2019,
  title = {Experimental and {{Theoretical Study}} of {{Thermodynamic Effects}} in a {{Quantum Annealer}}},
  author = {Kadowaki, Tadashi and Ohzeki, Masayuki},
  year = 2019,
  month = jun,
  journal = {Journal of the Physical Society of Japan},
  volume = {88},
  number = {6},
  pages = {061008},
  publisher = {The Physical Society of Japan},
  issn = {0031-9015},
  doi = {10.7566/JPSJ.88.061008},
  urldate = {2022-12-09}
}

@article{konz2019,
  title = {Uncertain Fate of Fair Sampling in Quantum Annealing},
  author = {K{\"o}nz, Mario S. and Mazzola, Guglielmo and Ochoa, Andrew J. and Katzgraber, Helmut G. and Troyer, Matthias},
  year = 2019,
  month = sep,
  journal = {Physical Review A},
  volume = {100},
  number = {3},
  pages = {030303},
  publisher = {American Physical Society},
  doi = {10.1103/PhysRevA.100.030303},
  urldate = {2021-03-16}
}

@article{mandra2017,
  title = {Exponentially {{Biased Ground-State Sampling}} of {{Quantum Annealing Machines}} with {{Transverse-Field Driving Hamiltonians}}},
  author = {Mandr{\`a}, Salvatore and Zhu, Zheng and Katzgraber, Helmut G.},
  year = 2017,
  month = feb,
  journal = {Physical Review Letters},
  volume = {118},
  number = {7},
  pages = {070502},
  publisher = {American Physical Society},
  doi = {10.1103/PhysRevLett.118.070502},
  urldate = {2021-01-27}
}

@article{marshall2020,
  title = {Perils of Embedding for Sampling Problems},
  author = {Marshall, Jeffrey and Di Gioacchino, Andrea and Rieffel, Eleanor G.},
  year = 2020,
  month = apr,
  journal = {Physical Review Research},
  volume = {2},
  number = {2},
  pages = {023020},
  publisher = {American Physical Society},
  doi = {10.1103/PhysRevResearch.2.023020},
  urldate = {2023-09-16}
}

@article{marshall2021,
  title = {Perils of {{Embedding}} for {{Quantum Sampling}}},
  author = {Marshall, Jeffrey and Mossi, Gianni and Rieffel, Eleanor G.},
  year = 2021,
  month = mar,
  journal = {arXiv:2103.07036 [quant-ph]},
  eprint = {2103.07036},
  primaryclass = {quant-ph},
  urldate = {2021-03-15},
  archiveprefix = {arXiv}
}

@article{matsuda2009,
  title = {Ground-State Statistics from Annealing Algorithms: Quantum versus Classical Approaches},
  shorttitle = {Ground-State Statistics from Annealing Algorithms},
  author = {Matsuda, Yoshiki and Nishimori, Hidetoshi and Katzgraber, Helmut G.},
  year = 2009,
  month = jul,
  journal = {New Journal of Physics},
  volume = {11},
  number = {7},
  pages = {073021},
  publisher = {IOP Publishing},
  issn = {1367-2630},
  doi = {10.1088/1367-2630/11/7/073021},
  urldate = {2021-01-27},
  langid = {english}
}

@article{neukart2017,
  title = {Traffic {{Flow Optimization Using}} a {{Quantum Annealer}}},
  author = {Neukart, Florian and Compostella, Gabriele and Seidel, Christian and {von Dollen}, David and Yarkoni, Sheir and Parney, Bob},
  year = 2017,
  month = dec,
  journal = {Frontiers in ICT},
  volume = {4},
  publisher = {Frontiers},
  issn = {2297-198X},
  doi = {10.3389/fict.2017.00029},
  urldate = {2025-08-01},
  langid = {english}
}

@article{ohzeki2018,
  title = {Optimization of Neural Networks via Finite-Value Quantum Fluctuations},
  shorttitle = {有限値量子ゆらぎを利用したニューラルネットワークの最適化},
  author = {Ohzeki, Masayuki and Okada, Shuntaro and Terabe, Masayoshi and Taguchi, Shinichiro},
  year = 2018,
  month = jul,
  journal = {Scientific Reports},
  volume = {8},
  number = {1},
  pages = {9950},
  issn = {2045-2322},
  doi = {10.1038/s41598-018-28212-4},
  urldate = {2023-01-26},
  langid = {english}
}

@misc{ohzeki2018a,
  title = {Control of Automated Guided Vehicles without Collision by Quantum Annealer and Digital Devices},
  author = {Ohzeki, Masayuki and Miki, Akira and Miyama, Masamichi J. and Terabe, Masayoshi},
  year = 2018,
  month = dec,
  journal = {arXiv.org},
  urldate = {2024-07-23},
  howpublished = {https://arxiv.org/abs/1812.01532v2},
  langid = {english}
}

@misc{pelofske2023,
  title = {Comparing {{Three Generations}} of {{D-Wave Quantum Annealers}} for {{Minor Embedded Combinatorial Optimization Problems}}},
  author = {Pelofske, Elijah},
  year = 2023,
  month = jan,
  number = {arXiv:2301.03009},
  eprint = {2301.03009},
  primaryclass = {quant-ph},
  publisher = {arXiv},
  doi = {10.48550/arXiv.2301.03009},
  urldate = {2023-02-25},
  archiveprefix = {arXiv}
}

@inproceedings{rosenberg2015,
  title = {Solving the Optimal Trading Trajectory Problem Using a Quantum Annealer},
  booktitle = {Proceedings of the 8th {{Workshop}} on {{High Performance Computational Finance}}},
  author = {Rosenberg, Gili and Haghnegahdar, Poya and Goddard, Phil and Carr, Peter and Wu, Kesheng and {de Prado}, Marcos L{\'o}pez},
  year = 2015,
  month = nov,
  series = {{{WHPCF}} '15},
  pages = {1--7},
  publisher = {Association for Computing Machinery},
  address = {New York, NY, USA},
  doi = {10.1145/2830556.2830563},
  urldate = {2025-08-01},
  isbn = {978-1-4503-4015-1}
}

@article{sieberer2018,
  title = {Programmable Superpositions of {{Ising}} Configurations},
  author = {Sieberer, Lukas M. and Lechner, Wolfgang},
  year = 2018,
  month = may,
  journal = {Physical Review A},
  volume = {97},
  number = {5},
  pages = {052329},
  publisher = {American Physical Society},
  doi = {10.1103/PhysRevA.97.052329},
  urldate = {2020-11-16},
  langid = {american}
}

@article{stollenwerk2020a,
  title = {Quantum {{Annealing Applied}} to {{De-Conflicting Optimal Trajectories}} for {{Air Traffic Management}}},
  author = {Stollenwerk, T. and O'Gorman, B. and Venturelli, D. and Mandr{\`a}, S. and Rodionova, O. and Ng, H. and Sridhar, B. and Rieffel, E. G. and Biswas, R.},
  year = 2020,
  month = jan,
  journal = {IEEE Transactions on Intelligent Transportation Systems},
  volume = {21},
  number = {1},
  pages = {285--297},
  issn = {1558-0016},
  doi = {10.1109/TITS.2019.2891235}
}

@article{suzuki2005,
  title = {Residual {{Energies}} after {{Slow Quantum Annealing}}},
  author = {Suzuki, Sei and Okada, Masato},
  year = 2005,
  month = jun,
  journal = {Journal of the Physical Society of Japan},
  volume = {74},
  number = {6},
  pages = {1649--1652},
  publisher = {The Physical Society of Japan},
  issn = {0031-9015},
  doi = {10.1143/JPSJ.74.1649},
  urldate = {2025-07-29}
}

@article{venturelli2016,
  title = {Quantum {{Annealing Implementation}} of {{Job-Shop Scheduling}}},
  author = {Venturelli, Davide and Marchand, Dominic J. J. and Rojo, Galo},
  year = 2016,
  month = oct,
  journal = {arXiv:1506.08479 [quant-ph]},
  eprint = {1506.08479},
  primaryclass = {quant-ph},
  urldate = {2021-04-05},
  archiveprefix = {arXiv}
}

@article{vinci2015,
  title = {Quantum Annealing Correction with Minor Embedding},
  author = {Vinci, Walter and Albash, Tameem and {Paz-Silva}, Gerardo and Hen, Itay and Lidar, Daniel A.},
  year = 2015,
  month = oct,
  journal = {Physical Review A},
  volume = {92},
  number = {4},
  pages = {042310},
  publisher = {American Physical Society},
  doi = {10.1103/PhysRevA.92.042310},
  urldate = {2024-02-02}
}

@article{weaver2012,
  title = {Satisfiability-Based {{Set Membership Filters}}},
  author = {Weaver, Sean A. and Ray, Katrina J. and Marek, Victor W. and Mayer, Andrew J. and Walker, Alden K.},
  year = 2012,
  month = jan,
  journal = {Journal on Satisfiability, Boolean Modeling and Computation},
  volume = {8},
  number = {3-4},
  pages = {129--148},
  publisher = {IOS Press},
  doi = {10.3233/SAT190095},
  urldate = {2021-03-07},
  langid = {english}
}

\end{document}